# Size Effect of Local Current-Voltage Characteristics of MX₂ Nanoflakes: Local Density of States Reconstruction from Scanning Tunneling Microscopy Experiments

Anna N. Morozovska[1*], Hanna V. Shevliakova[1,2], Yaroslava Yu. Lopatina[1], Mykola Yelisieiev[3], Galina I. Dovbeshko[1], Marina V. Olenchuk[1], G.S. Svechnikov[2], Sergei V. Kalinin[4,†], Yunseok Kim[5‡], and Eugene A. Eliseev[6§]

[1]*Institute of Physics, National Academy of Sciences of Ukraine,*
*46, pr. Nauky, 03028 Kyiv, Ukraine*

[2] *Department of Microelectronics, National Technical University of Ukraine "Igor Sikorsky Kyiv Polytechnic Institute", Kyiv, Ukraine*

[3] *Taras Shevchenko National University of Kyiv, Volodymyrska street 64, Kyiv, 01601, Ukraine*

[4] *The Center for Nanophase Materials Sciences, Oak Ridge National Laboratory, Oak Ridge, TN 37922*

[5] *School of Advanced Materials Science and Engineering, Sungkyunkwan University (SKKU), Suwon 16419, Republic of Korea*

[6] *Institute for Problems of Materials Science, National Academy of Sciences of Ukraine, Krjijanovskogo 3, 03142 Kyiv, Ukraine*

## Abstract

Local current-voltage characteristics for low-dimensional transition metal dichalcogenides (LD-TMD), as well as the reconstruction of their local density of states (LDOS) from scanning tunneling microscopy (STM) experiments is of fundamental interest and can be useful for advanced applications. Most of existing models are either hardly applicable for the LD-TMD of complex shape (e.g., those based on Simmons approach), or necessary for solving an ill-defined integral equation to deconvolute the unknown LDOS (e.g., those based on Tersoff approach). Using a serial expansion of Tersoff formulae, we propose a flexible method how to reconstruct the LDOS from local current-voltage characteristics measured in STM experiments. We established a set of key physical parameters, which characterize the tunneling current of a STM probe – sample contact and the sample LDOS expanded in Gaussian functions. Using a direct variational method coupled with a probabilistic analysis, we determine these parameters from the STM experiments for $MoS_2$ nanoflakes with different number of layers. The main result is the reconstruction of the LDOS in a relatively wide energy range around a Fermi level, which allows insight in the local band structure of LD-TMDs. The reconstructed LDOS reveals pronounced size effects for the single-layer, bi-layer and three-layer $MoS_2$ nanoflakes, which we relate with low dimensionality and strong bending/corrugation of the nanoflakes. We hope that the proposed elaboration of the Tersoff approach allowing LDOS reconstruction will be of urgent interest for quantitative description of STM experiments, as well as useful for the microscopic physical understanding of the surface, strain and bending contribution to LD-TMDs electronic properties.

---

[*] Corresponding author 1: anna.n.morozovska@gmail.com
[†] Corresponding author 2: sergei2@ornl.gov
[‡] Corresponding author 3: yunseokkim@skku.edu
[§] Corresponding author 4: eugene.a.eliseev@gmail.com

## I. Introduction

The tunneling effect is widely used in solid state physics for spectroscopy of electronic states. The method is based on the dependence of the tunneling current on the number of states that form the tunnel contact in semiconductors in the energy range from 0 to eV, which is counted from the Fermi level $E_F$, and V is the tunneling gap bias. For the scanning tunneling microscopy (**STM**) it is necessary to take into account the coordinate dependence of the local density of states (**LDOS**), which opens the possibility of the scanning tunneling spectroscopy (**STS**) with high spatial resolution [1].

Local current-voltage (**I-V**) characteristics for low-dimensional (**LD**) transition metal dichalcogenides (**TMD**), as well as the reconstruction of their LDOS from STM I-V curves is of fundamental interest and can be useful for advanced applications, primary because their semiconducting properties are versatile and tunable [2, 3]. For instance, the structural, polar and electronic properties of the low-dimensional $MX_2$ (M – metal Mo, V, W; X – chalcogen S, Se, Te) [4, 5] and Janus-compounds MXY (X, Y – chalcogens) [6, 7] are varying from non-polar to ferroelectric state, and from direct-band semiconductor to metallic conductivity [8, 9]. The studies of the LD $MoX_2$ local electrical conductivity attract permanent interest [10, 11, 12]; and the property is readily tunable. In particular, the structural and electronic properties of LD-$MoS_2$ may be different depending on the preparation conditions, concentration of impurities and defects [13, 14, 15, 16]. The local bending of MoX$_2$ nanoflakes can lead to the significant dependence of the free carrier concentration on the strain gradient [17, 18, 19].

Most of existing models for the determination of LDOS from local I-V curves are hardly applicable for the LD-TMD of complex shape, such as bended and/or corrugated nanolayers, nanoflakes and ultra-small nanoparticles. In particular, the one-dimensional tunneling current model, proposed by Simmons [20, 21], is a priori not applicable to nanoflakes due to the influence of strongly inhomogeneous "transverse" electric fields that can be created by bending flakes and other edge effects. In addition, this model is not applicable to atomically-sharp probes. More suitable is the Tersoff approach [22], which is widely used in STS and potentially allows to determine the LDOS of the studied substrate at Fermi level; also it can be adapted to modern atomically-sharp probes and substrate inhomogeneities [1, 23]. However, the Tersoff [22] approach requires solving an ill-defined integral equation to deconvolute the unknown LDOS in the finite energy range around the Fermi level.

Using a serial expansion of Tersoff formulae [22], in this work, we propose a method how to reconstruct the LDOS from the local I-V curves measured in STM experiments for MoS$_2$ nanoflakes with different number of layers. Below we explain how to reconstruct the LDOS in a relatively wide energy range around a Fermi level, and reveal pronounced size effects for the single-layer, bi-layer and three-layer MoS$_2$ nanoflakes.

The manuscript is structured as following. **Section II** contains the formulation of the problem for the LDOS reconstruction from the STM current and analytical expressions for the tunneling current derived for LDOS expanded in Gaussian functions. **Section III** is analysis of the local I-V curves measured by STM in a single-layer, bi-layer and three-layer MoS$_2$ nanoflakes, and reconstruct the LDOS from these curves. **Section IV** is a brief summary. Calculations details and auxiliary figures are listed in **Suppl. Mat**. [24].

### II. Modelling of the tunneling current and LDOS of MoS$_2$ nanoflakes

Typical STM images of MoS$_2$ nanoflakes on HOPG and surface topography along the AB direction are shown in **Figs. 1(a)-1(b)**. STM images of different parts of the samples were detected by MoS$_2$ nanoflakes of arbitrary shape and size.

It was shown in Ref.[25] that bulk MoS$_2$ is an indirect band semiconductor with a band gap of 1.2 eV, which begins in the $\Gamma$ point, and ends in the conductive band bottom, halfway between $\Gamma$ and K points. As the number of layers decreases, the main indirect band gap (from point $\Gamma$ to the middle between point $\Gamma$ and point K) increases due to a quantum constraint and becomes larger than the direct band gap located at point K in the case of a MoS$_2$ monolayer.

Within the *flat* monolayer, MoS$_2$ is transformed from a bulk non-indirect semiconductor into a two-dimensional straight-band semiconductor with a larger band gap. Indirect band gap of a bulk MoS$_2$ (1.2 eV) is replaced by direct band gap (1.9 eV) at point K of monolayer MoS$_2$. The local strains, large strain gradients, bending and especially corrugation can strongly affect the DOS of MoS$_2$ nanoflakes making them semi-metallic or metallic in 1T' phase [7, 8, 17-19]. To probe the influence of the strains, we use strongly deformed nanoflakes. The strong deformation of the studied nanoflakes is clearly visible from a typical SEM image in **Figs.1(c)**.

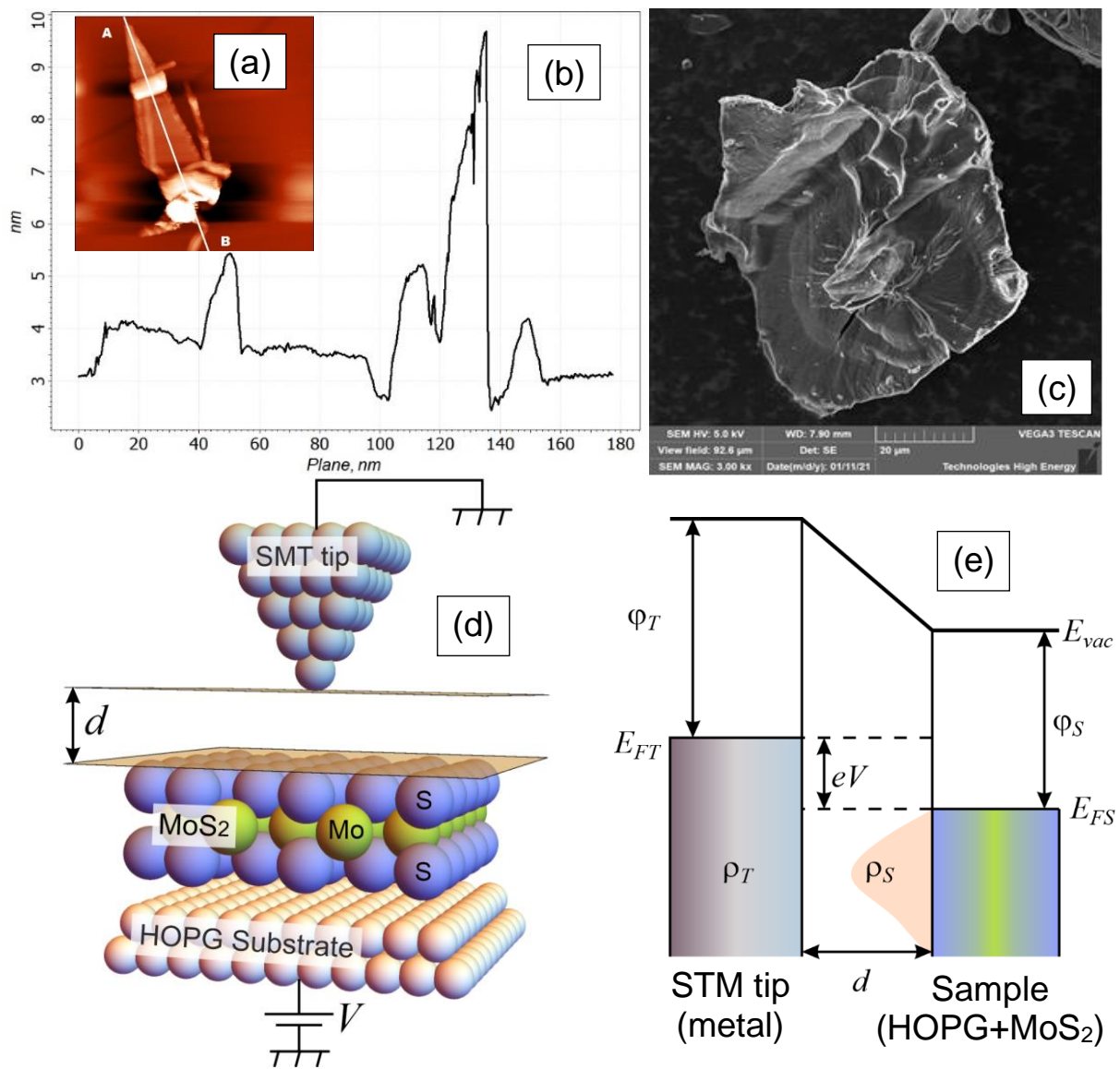

**Figure 1. (a)** "Sparking" and corrugated nanoflake placed between the STM tip and conducting HOPG substrate. **(b)** Surface topography along the direction AB. **(c)** SEM image of a corrugated single-layer MoS$_2$ nanoflake. Schematics of STM experiment **(d)** and zone scheme **(e)** used in our calculations.

Using multidimensional tunneling current models [26] and the perturbation theory, the tunneling current $I$ through the space between the sample and the probe can be written as

$$I(x,y) = \frac{2\pi e}{h} \sum_{\mu\nu} f(E_\mu) \left[1 - f(E_\nu + eV)\right] \left|M_{\mu\nu}(x,y)\right|^2 \delta(E_\mu - E_\nu), \quad (1a)$$

where $f(E) = \frac{1}{1 + exp\left(\frac{E - E_F}{k_B T}\right)}$ is the Fermi function, $E_F$ is the Fermi level, $eV$ is applied voltage, which may depend on the "transverse" surface relief. The matrix element $M_{\mu\nu}$ is

$$M_{\mu\nu} = -\frac{\hbar^2}{2m} \int \left(\psi_\mu^* \nabla \psi_\nu - \psi_\nu^* \nabla \psi_\mu\right) ds \quad (1b)$$

between the undisturbed states of the probe $\psi_\mu$ and the sample surface $\psi_\nu$, $E_\mu$ is the energy of the state $\psi_\mu$ without tunneling. Integration is carried out on any surface $S$ in the air/vacuum gap between the probe and the surface. According to the perturbation theory, these wave functions satisfy the stationary Schrödinger equation $-\frac{\hbar^2}{2m}\Delta\psi + e\left(V_S + V_T\right)\psi = E\,\psi$, where the potential $V_T$, which is created by the STM probe, is considered a small perturbation in comparison with the potential of the sample surface $V_S$. Summation in Eq. (1a) is conducted on all possible electronic states (see **Appendix A** for details).

When performing STM studies on semiconductor samples, the applied voltage V can reach (0.5 - 2) V, which is not small compared to $k_B T$ for room temperature (26 meV). At such voltages, the local densities of the probe and substrate states still can be introduced, but the point probe approximation, which is valid at very low voltages of ~(1-10)mV, cannot be used. Under these conditions, equation (1a) can be roughly written in the following form [27]:

$$I(\boldsymbol{r}, V) \cong \frac{4\pi e}{h} \int_{E_F}^{E_F + eV} \rho_T(E + eV)\, \tilde{\rho}_S(\boldsymbol{r}, E) T(\boldsymbol{r}, E, V) dE, \quad (2)$$

where $\rho_T(E + eV)$ is the LDOS related with the probe, $\tilde{\rho}_S(\boldsymbol{r}, E)$ is the LDOS related with the sample in the point $\boldsymbol{r}$, and $T(\boldsymbol{r}, E, V) = |M|^2$ is the local probability of the electron tunneling. Note that if LDOS has a singularity at the Fermi energy $E = E_F$, the value of $I(\boldsymbol{r}, 0)$ is nonzero and determined by the surface states [27], work function difference and/or Schottky barriers. For the purposes of correct fitting, let us separate $I(\boldsymbol{r}, 0) \equiv I_S$ and introduce the "regular" part of the sample LDOS denoted as $\rho_S(\boldsymbol{r}, E)$.

Within Wentzel–Kramers–Brillouin (**WKB**) approximation, the value of the barrier transparency coefficient $T(\boldsymbol{r}, E, eV)$ is given by expression [1]:

$$T(\boldsymbol{r}, E, V) = \exp[-2\kappa d] \cong \exp\left[-2d\sqrt{\frac{2me}{\hbar^2}\left(\frac{\varphi_T + \varphi_S}{2} - E + \frac{V}{2}\right)}\right], \quad (3a)$$

where $d$ is probe-sample distance, which can also vary during the scanning (e.g., in DC mode), and $\kappa$ is a damping constant. Formulae (3a) is based on the assumption that only the s-functions of the probe make a major contribution to its LDOS near the Fermi level, $\rho_T(E_F)$, that is proportional to $|\psi_\mu|^2 \sim exp(-2\kappa d)$. The damping constant is given by the following expression:

$$\kappa = \sqrt{\frac{2me}{\hbar^2}\left(\frac{\varphi_T+\varphi_S}{2}-E+\frac{V}{2}\right)}, \qquad (3b)$$

where $\varphi_T$ and $\varphi_S$ are the work functions of the electron at the surface of the probe and at the sample surface, respectively. Hereinafter we measure them in a voltage unit.

If the sample LDOS, $\rho_S(\boldsymbol{r},E)$, is dominated in the spectrum, and the probe states $\rho_T$ are stable, the equation (2) can be simplified as:

$$I(\boldsymbol{r},V) \cong I_S + \frac{4\pi}{\hbar}\rho_T\int_0^V \rho_S(\boldsymbol{r},E+E_F)\exp\left[-2d\sqrt{\frac{2me}{\hbar^2}\left(\frac{\varphi_T+\varphi_S}{2}-E_F-E+\frac{V}{2}\right)}\right]dE, \qquad (4)$$

The "residual" current $I_S$ is related with the LDOS peculiarity at $E = E_F$ in Eq.(2). All energies in Eq.(4) is written in a voltage unit, and the condition $\frac{\varphi_T+\varphi_S}{2}-E_F-E+\frac{V}{2} \geq 0$ should be valid.

Formula (4) is suitable for the deconvolution of the sample LDOS with STS maps. Since a tunneling current is measured with a significant error, it makes sense to deconvolve it using the direct variation method, similarly as it was done for piezoresponse force spectroscopy [28].

The tunneling current values $I_n(\boldsymbol{r})$ are measured for the voltages $V_n$ where $n = 1, \dots N$. For each scan point **r**, we need to find the parameters of the transparency coefficient $T(\boldsymbol{r},E,eV)$ and the DOS $\rho_S(\boldsymbol{r},E)$, which minimize the following functional:

$$Q(\boldsymbol{r}) = \frac{1}{N}\sum_{n=1}^{N}[I_n(\boldsymbol{r}) - I(\boldsymbol{r},V_n)]^2 . \qquad (5)$$

As it can be seen from Eq.(4), the parameters are the probe-sample distance $d$, the voltage shift $V_S$, sum of work functions $\phi = \frac{\varphi_T+\varphi_S}{2}$, and the sample LDOS $\rho_S(\boldsymbol{r},E)$, which form of is a priori unknown, but may significantly depends on the number of MoS$_2$ layers and the local electric fields, which "distort" it. Actually, the selection of the form of this function and its parameters is the main hypothesis of the model. We need to go through several such hypotheses and choose the most appropriate. Below we use the sum of two Gaussian functions, which can model LDOS of intrinsic, p-type or n-type semiconductor, semimetal or metal, namely:

$$\rho_S(\boldsymbol{r},E) = g_e\exp\left[-\frac{(E-E_F-E_e)^2}{\Delta_e^2}\right] + g_p\exp\left[-\frac{(E-E_F+E_p)^2}{\Delta_p^2}\right]. \qquad (6a)$$

The LDOS in Eq.(6) contains six fitting parameters: number of electron and hole states, $g_e$ and $g_p$, the maximum positions of corresponding DOS, $E_e$ and $E_p$, and their dispersions, $\Delta_e$ and $\Delta_p$, respectively. The Fermi energy $E_F$ is determined by the band structure of the system "tip-gap-sample-substrate". For the impurity-free case $g_e = g_p = g$, $\Delta_e = \Delta_p = \Delta$ and $E_e = E_p$. The situation of $g_e \gg g_p$ corresponds to the n-type doping, and $g_e \ll g_p$ is valid for the p-type doping. Below we omit the argument "$\boldsymbol{r}$" in $\rho_s(\boldsymbol{r}, E)$ for the sake of simplicity.

For flat and weakly deformed areas of multilayer MoS$_2$ nanoflakes, their LDOS may be close to a proper semiconductor

$$g_e \cong g_p, \;\; \Delta_e = \Delta_p = \Delta, \;\; E_e = E_p = E_s, \;\; |E_s| \geq 3\Delta \quad \text{(hypothesis 1)}, \quad (6b)$$

Using the results obtained by the authors of [29], it can be assumed that, for strongly bended sections of corrugated MoS$_2$ nanoflakes, the LDOS can be close to a semi-metal or even metal in 1T' phase. This means that the electron and hole LDOS are overlapped, and so we can assume that

$$g_e \sim g_p, \;\; \Delta_e = \Delta_p = \Delta, \;\; E_e = E_p = E_s, \;\; \Delta \leq |E_s| \leq 3\Delta \;\; \text{(hypothesis 2)}, \quad (6c)$$

For a metallic n-type MoS$_2$ and/or electron-conducting substrate, the LDOS is maximal at the Fermi energy

$$g_e \gg g_p, \;\; \Delta_e = \Delta_p = \Delta, \;\; E_e = E_p = E_s, \;\; |E_s| < \Delta, \quad \text{(hypothesis 3)}. \quad (6d)$$

The p-type case is given by Eq.(6d) with $g_e \ll g_p$.

Sketches of the LDOS calculated from Eq. (6) are shown in **Fig. 2**.

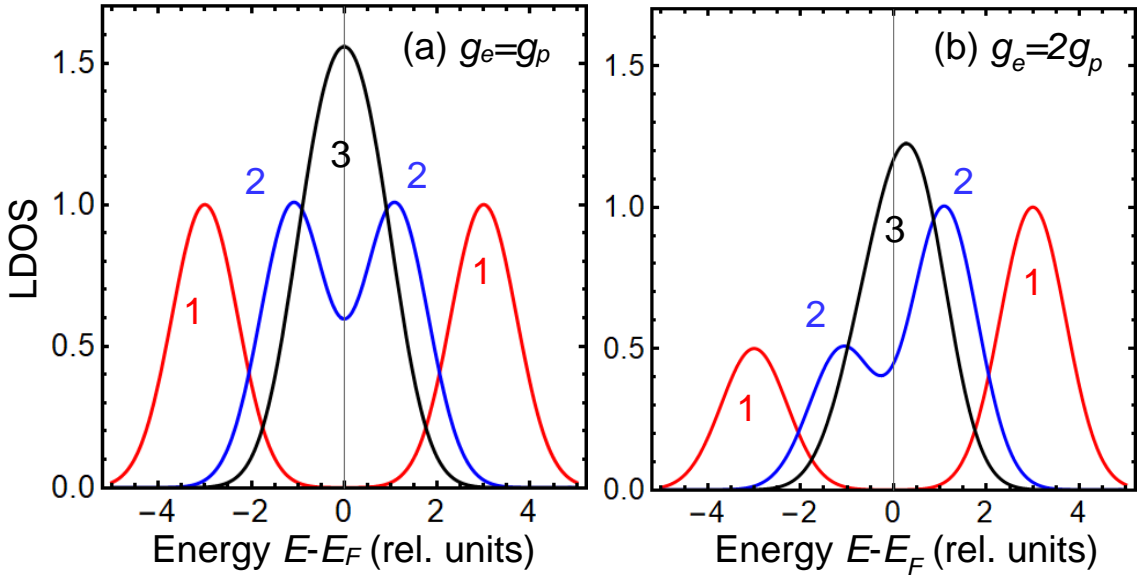

**Figure 2**. Red, blue and black curves illustrate LDOS for hypothesis 1, 2 and 3, respectively, depending on the parameters $\Delta_e = \Delta_p = \Delta$, $E_e = E_p = E_s$, $g_e = g_p$ **(a)** and $g_e = 2g_p$ **(b).**

Priory to apply the hypothesis 1-3, let us expand the tunneling current of Eq. (4) in series on the voltage powers:

$$I(V) \cong I_S + \sum_{i=1}^{n} \sigma_i V^i \equiv \sum_{i=1}^{n} \tilde{\sigma}_i (V - V_S)^i, \tag{7a}$$

where $n$ is an integer, and $V_S$ is a voltage shift determined by the offset current $I_S$ and vise a versa. The first five expansion coefficients $\sigma_i$ are given by expressions:

$$\sigma_1 = \rho e^{-\sqrt{Z\phi}} \rho_S(E_F), \tag{7b}$$

$$\sigma_2 = \frac{\rho}{2} e^{-\sqrt{Z\phi}} \frac{\partial \rho_S(E)}{\partial E}\Big|_{E \to E_F}, \tag{7c}$$

$$\sigma_3 = \frac{\rho}{6} e^{-\sqrt{Z\phi}} \left[ \frac{Z\phi + \sqrt{Z\phi}}{16\phi^2} \rho_S(E) + \frac{1}{4}\sqrt{\frac{Z}{\phi}} \frac{\partial \rho_S(E)}{\partial E} + \frac{\partial^2 \rho_S(E)}{\partial E^2} \right]\Big|_{E \to E_F}, \tag{7d}$$

$$\sigma_4 = \frac{\rho}{24} e^{-\sqrt{Z\phi}} \left[ \frac{Z\phi + \sqrt{Z\phi}}{8\phi^2} \frac{\partial \rho_S(E)}{\partial E} + \frac{1}{2}\sqrt{\frac{Z}{\phi}} \frac{\partial^2 \rho_S(E)}{\partial E^2} + \frac{\partial^3 \rho_S(E)}{\partial E^3} \right]\Big|_{E \to E_F}, \tag{7e}$$

$$\sigma_5 = \frac{\rho}{120} e^{-\sqrt{Z\phi}} \left[ \frac{15Z\phi + Z^2\phi^2 + 15\sqrt{Z\phi} + 6(Z\phi)^{3/2}}{256\phi^4} \rho_S(E) + \frac{3Z\phi + 3\sqrt{Z\phi} + (Z\phi)^{3/2}}{32\phi^3} \frac{\partial \rho_S(E)}{\partial E} + \right.$$

$$\left. \frac{Z\phi + \sqrt{Z\phi}}{4\phi^2} \frac{\partial^2 \rho_S(E)}{\partial E^2} + \frac{3\sqrt{Z\phi}}{4\phi} \frac{\partial^3 \rho_S(E)}{\partial E^3} + \frac{\partial^4 \rho_S(E)}{\partial E^4} \right]\Big|_{E \to E_F}, \tag{7f}$$

where we introduced the factor $Z = \frac{8em}{h^2} d^2$ proportional to the second power of the tip-sample separation $d$, the barrier height $\phi = \frac{\varphi_T + \varphi_S}{2} - E_F$, and the tip LDOS magnitude $\rho = \frac{4\pi}{h} \rho_T$. More complex expressions for $\tilde{\sigma}_i$ are considered in **Appendix B**.

Equations (7) are valid for arbitrary LDOS. Here the first expansion coefficient $\sigma_1$ contains the information about the sample LDOS at the local Fermi level, $\rho_S(E_F)$, the second and other coefficients contain the mixed information about the LDOS and its derivatives, $\frac{\partial^i \rho_S(E)}{\partial E^i}$ at $E \to E_F$.

The system (7) written for $n$ coefficients $\sigma_n$ contains $n$ unknown characteristics of the sample LDOS, $\rho_S(E_F)$, $\frac{\partial \rho_S(E_F)}{\partial E}$,..., $\frac{\partial^{n-1} \rho_S(E_F)}{\partial E^{n-1}}$, and three unknown parameters of the tip, tunneling gap and contact, $\rho$, $Z$ and $\phi$, which are included as combinations of $\rho e^{-\sqrt{Z\phi}}$, $Z\phi$ and $\frac{1}{\phi}$. Since the system of $n$ equations contain $n + 3$ unknown variables, it is ill-defined for the unique reconstruction of the Tailor series, $\rho_S(E) = \rho_S(E_F) + \sum_{n=1}^{\infty} \frac{\partial^n \rho_S(E_F)}{\partial E^n} \frac{(E - E_F)^n}{n!}$, for arbitrary $\rho_S(E)$. To solve approximately the ill-defined problem, we need to use a hypothesis for $\rho_S(E)$ allowing to decrease significantly the number of unknown parameters in $\rho_S(E)$.

For the LDOS from Eq. (6a), the first expansion coefficients $\sigma_i$ are given by expressions:

$$\sigma_1 = \rho(g_e + g_p)\exp\left[-\sqrt{Z\phi} - \frac{E_s^2}{\Delta^2}\right], \tag{8a}$$

$$\sigma_2 = \rho(g_e - g_p)\frac{E_s}{\Delta^2}\exp\left[-\sqrt{Z\phi} - \frac{E_s^2}{\Delta^2}\right], \tag{8b}$$

$$\sigma_3 = \frac{\rho}{6}\exp\left[-\sqrt{Z\phi} - \frac{E_s^2}{\Delta^2}\right]\left[\frac{Z\phi + \sqrt{Z\phi}}{16\phi^2}(g_e + g_p) + \frac{1}{2}\sqrt{\frac{Z}{\phi}}(g_e - g_p)\frac{E_s}{\Delta^2} + 2(g_e + g_p)\frac{2E_s^2 - \Delta^2}{\Delta^4}\right], \tag{8c}$$

$$\sigma_4 = \frac{\rho}{24}\exp\left[-\sqrt{Z\phi} - \frac{E_s^2}{\Delta^2}\right]\left[\frac{Z\phi + \sqrt{Z\phi}}{4\phi^2}(g_e - g_p)\frac{E_s}{\Delta^2} + \sqrt{\frac{Z}{\phi}}\frac{2E_s^2 - \Delta^2}{2\Delta^4}(g_e + g_p) + \right.$$
$$\left. 4\frac{2E_s^2 - 3\Delta^2}{\Delta^4}\frac{E_s}{\Delta^2}(g_e - g_p)\right]. \tag{8d}$$

$$\sigma_5 = \frac{\rho}{120}\exp\left[-\sqrt{Z\phi} - \frac{E_s^2}{\Delta^2}\right]\left[\frac{15Z\phi + Z^2\phi^2 + 15\sqrt{Z\phi} + 6(Z\phi)^{3/2}}{256\phi^4}(g_e + g_p) + \right.$$
$$\frac{3Z\phi + 3\sqrt{Z\phi} + (Z\phi)^{3/2}}{16\phi^3}(g_e - g_p)\frac{E_s}{\Delta^2} + \frac{Z\phi + \sqrt{Z\phi}}{2\phi^2}\frac{2E_s^2 - \Delta^2}{\Delta^4}(g_e + g_p) + 3\sqrt{\frac{Z}{\phi}}\frac{2E_s^2 - 3\Delta^2}{\Delta^4}(g_e - $$
$$g_p)\frac{E_s}{\Delta^2} + 4\frac{4E_s^2(E_s^2 - 3\Delta^2) + 3\Delta^4}{\Delta^8}(g_e + g_p)\right]. \tag{8e}$$

The system (8) written for $n$ coefficients $\sigma_n$ contains 4 unknown characteristics of the sample LDOS, $g_e$, $g_p$, $E_s$ and $\Delta^2$, and three unknown parameters of the tip-sample contact, $\rho$, $Z$ and $\phi$. So, rigorously we need seven experimental values of $\sigma_1 - \sigma_7$ for fitting. However, the accuracy of the polynomial fitting of the local I-V curves in the range of small voltages may not allow to determine so many coefficients $\sigma_i$ without significant noise-related error. For instance, for the STM data considered in the next section only the first 4-5 coefficients $\sigma_i$ can be reliably extracted from the data. This noise-related problem subjects us to use probabilistic analysis for the determination of the unknown LDOS.

By equating the expressions (8) for the coefficients $\sigma_i^{(t)}$ determined from the fitting of STM current $\sigma_i^{exp}$, we can extract the unknown parameters of the nanoflake effective LDOS. Using the probability analysis (in fact Bayesian formulae) we obtain the following system of equations for the determination of $P_j$:

$$\sigma_i^{exp} = \sum_{j=1}^{3} P_j\sigma_i^{(j)}, \qquad \sum_{j=1}^{3} P_j = 1, \quad 0 \le P_j \le 1. \tag{9}$$

The probabilities $P_j$ of a priory $j$-th hypothesis realization along with corresponding LDOS parameters can be determined. Note that the Bayesian analysis has been successfully used for atomically resolved STEM data [30].

### III. Determination of the DOS properties from STM experiments

Measured by STM, local I-V curves of a HOPG substrate without nanoflakes are shown in **Fig. 3(a)**. Local I-V curves for a single-layer, bilayer and three-layer $MoS_2$ nanoflakes of thickness $h \approx 0.63$ nm, $h \approx 1.21$ nm and $h \approx 2.03$ nm are shown in **Figs. 3(b)**, **3(c)** and **3(d)**, respectively.

These local I – V curves were measured in different moments of time at the nanoflake points marked with a white circle in the insets in **Fig. 3**. Also, we collected 20 curves in the forward (red symbols) and backward (blue symbols) directions. The variability of the local I– V curves and typical jumps on them indicate the randomness of a particular path of tunneling current associated with local inhomogeneity of electrical conductivity near the surface of $MoS_2$ nanoflakes, the evolution of which is most likely caused by the changes in their shape and dynamic deformation during the current flow. The charge in the inhomogeneity of electrical conductivity is bipolar, as can be seen from the local I – V curves.

It is seen from **Fig. 3**, that the forward and backward local I-V curves are well-fitted with Eq. (8) (black curves). From the fitting we determined the tunneling current offset $I_s$, the voltage shift $V_S$ and the expansion coefficients $\sigma_i$. We are especially interested in the dependences of the fitting parameters on the thickness $h$ of $MoS_2$ nanoflakes, which can be interpreted as a size effect. Below we use that $h = Nh_S$, where $N = 0, 1, 2, 3$ and $h_S \approx 0.63$ nm is the thickness of s single-layer $MoS_2$.

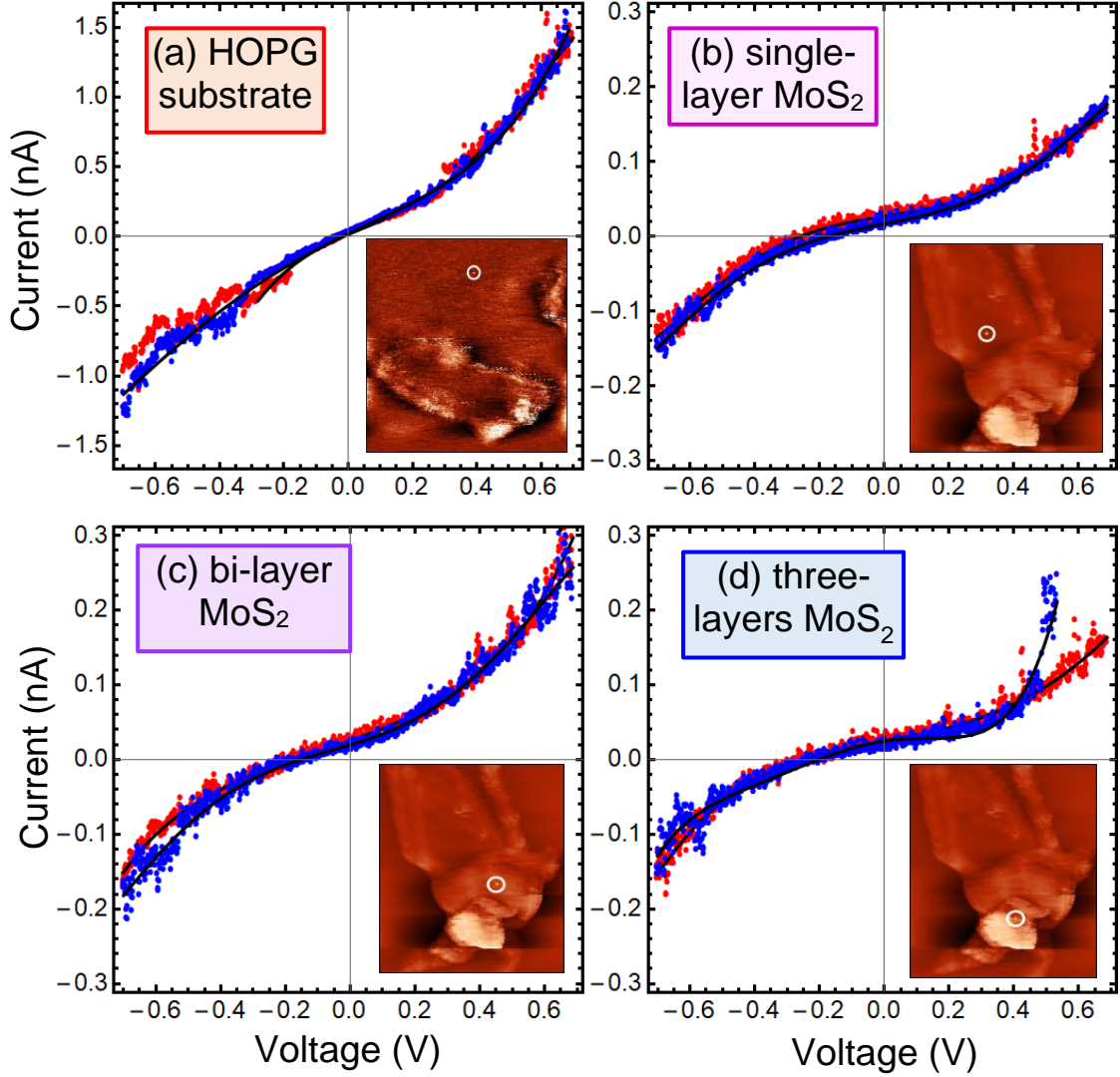

**Figure 3.** Local I-V curves measured in the forward (red symbols) and backward (blue symbols) directions, and their fitting with Eq.(8) (black curves). The I-V curves were measured by STM in the middle of white circle of HOPG substrate (**a**), single-layer, bi-layer and three-layer MoS$_2$ nanoflakes of thickness $h \approx 0.63$ nm (**b**), $h \approx 1.21$ nm (**c**) and $h \approx 2.03$ nm (**d**), respectively.

The dependence of the offset $I_s$, voltage shift $V_S$ and fitting parameters $\sigma_i$ on the number $N$ of MoS$_2$ layers are shown in **Fig. 4.** The residual value of tunneling current $I_s(N)$ decreases in more than two times with $N$ increase from 0 to 3 for the backward run, and increases less than in 2 times for the forward run [compare red and blue curves in **Fig. 4(a)**]. The dependences $I_s(N)$ are nonmonotonic and slightly different for the forward and backward direction ("run") of the I-V curves. The highest error bars are for $N = 0$ corresponding to the HOPG substrate, while the errors are rather small for $N = 1 - 3$.

The voltage shift $V_s$ is always negative and its absolute value significantly decreases from very small values about -0.01 V to -0.25 V with $N$ increase from 0 to 1, then $|V_s|$ decreases for $N = 2$ and then increases again for $N = 3$ [see **Fig. 4(b)**]. The nonmonotonic dependences $V_s(N)$ are similar for both backward and forward runs. Allowing for the error bars for $N = 0$, size dependences of the odd coefficients $\sigma_1(N)$ and $\sigma_3(N)$ are quasi-monotonic; they decrease from relatively high positive values (1 nA/V and 6 nA/V$^3$, respectively) to very small positive values with $N$ increase from 0 to 1 [compare **Figs. 4(c)** and **4(e)**]. Allowing for the error bars for $N = 0$, size dependences of the even coefficients $\sigma_2(N)$ and $\sigma_4(N)$ are also quasi-monotonic, but their signs are different; they decrease from relatively high negative (or positive) values to very small negative (or positive) values with $N$ increase from 0 to 1 [compare **Figs. 4(d)** and **4(f)**]. The functional form and values of the dependences $\sigma_i(N)$ are rather similar for both backward and forward runs. It is important to mention that the higher coefficients $\sigma_6(N)$ and $\sigma_7(N)$ appeared zero with accuracy of experimental error. So, their formal inclusion in order to determine LDOS parameters are not grounded.

Since the size dependencies of the fitting parameters, $V_S(N)$ and $\sigma_i(N)$ are regular, their functional form and values are close for both backward and forward runs of the local I-V curves, we have an opportunity to determine accurately ("deconvolute") the parameters of HOPG and MoS$_2$ nanoflakes LDOS given by e.g., Eq.(6a) using the data shown in **Fig. 4.**

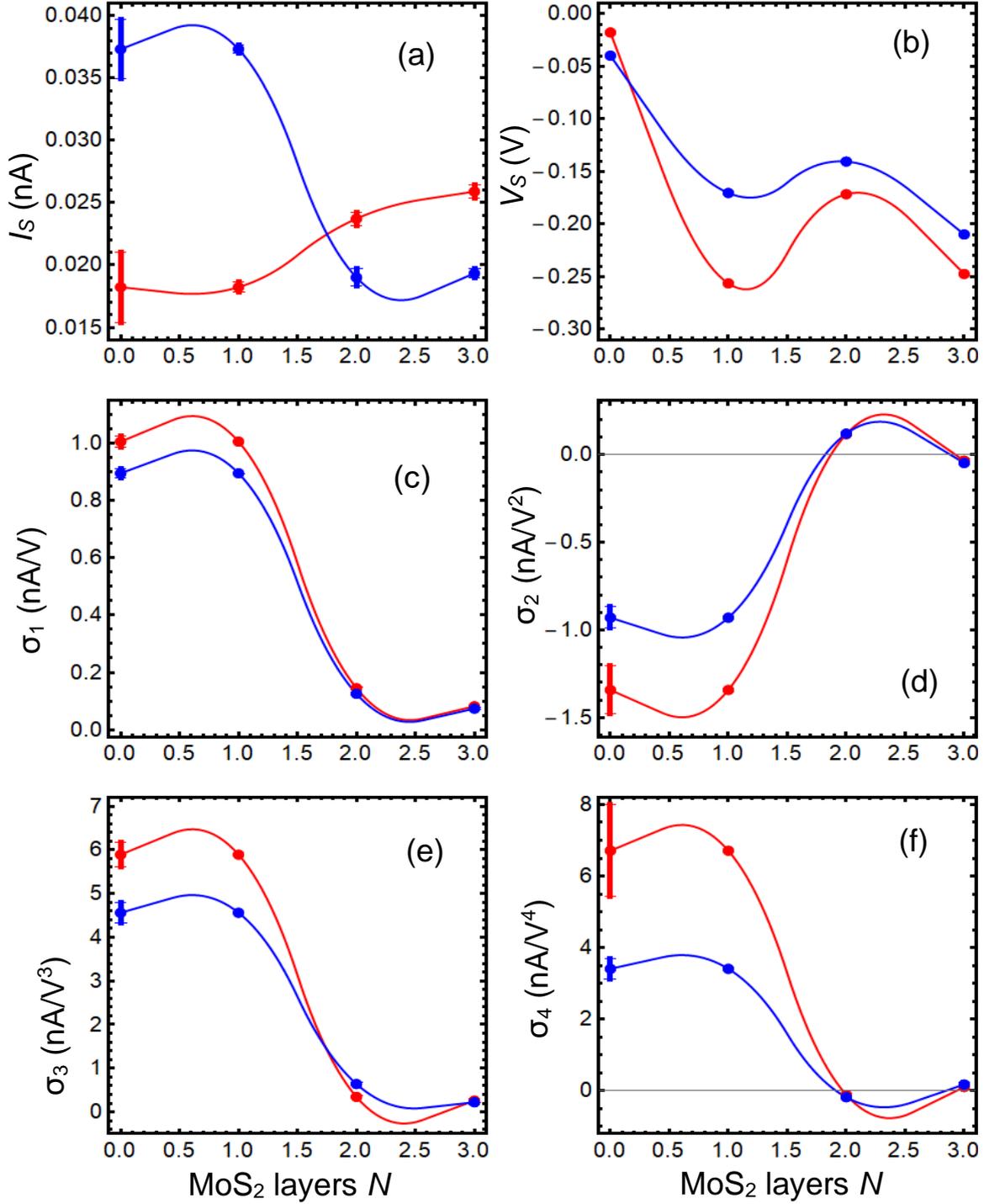

**Figure 4.** The dependence of the offset $I_s$ (**a**), voltage shift $V_S$ (**b**), and fitting parameters $\sigma_i$ (**c**)-(**f**), on the number $N$ of $MoS_2$ layers (0, 1, 2, and 3) extracted for the forward (red symbols) and backward (blue symbols) local I-V curves of $MoS_2$ nanoflakes on HOPG substrate, shown in **Fig. 3.** Solid red and blue curves are spline-interpolations.

To make the deconvolution procedure more reliable, hereinafter we introduce positive dimensionless parameters, which determine the Gaussian-type LDOS $\rho_s(\boldsymbol{r}, E)$ in Eq.(6a) and the barrier transparency coefficient $T(\boldsymbol{r}, E, V)$ in Eq.(3a):

$$g = \frac{g_e - g_p}{g_e + g_p}, \quad f = \frac{\Delta}{|\phi|}, \quad \varepsilon = \frac{E_s}{\Delta}, \quad \delta = \frac{\Delta}{E_g}, \quad R = \sqrt{Z\phi}, \qquad (10)$$

where the dispersion $\Delta$ is dimensionalized on the bulk band gap of $MoS_2$, $E_g = 1.2\text{eV}$.

Parameters (10) have a simple physical sense. The parameter $g$ determines the ratio of the electron to hole states and varies from -1 to 1. The parameter $f$ is the ratio of the sample LDOS halfwidth to the energy factor $|\phi|$ in the transparency $T(\boldsymbol{r}, E, V)$. The parameter $\varepsilon$ is the ratio of the electron/hole LDOS maximum to its halfwidth. The parameter $\delta$ is the ratio of the electron/hole LDOS halfwidth to the bulk band gap of $MoS_2$. The parameter $R$ is proportional to the STM tip-surface separation $d$. Since the parameter $R$ determines the exponential factor in $T(\boldsymbol{r}, E, V)$, its values cannot be very small, otherwise the tunneling current $I(\boldsymbol{r}, V)$ vanishes. On the other hand, it cannot be very small, because STM operates in the tunneling regime. The range $0.5 - 5$ seems optimal. Using Eqs.(10) one could rewrite the system of equations (8) in the dimensionless variables as listed in **Appendix C**, and then determine the LDOS parameters numerically.

The dependences of the dimensionless parameters $g$, $f$, $\varepsilon$ and $\delta$ on the parameter $R$ calculated for HOPG, and 1, 2, and 3 $MoS_2$ are shown in **Fig. 5**. Left column corresponds to the forward, and right column corresponds to the backward run of local I-V curves of a $MoS_2$ on HOPG substrate, shown in **Fig. 3.** The dependences $g(R)$, $f(R)$, $\varepsilon(R)$ and $\delta(R)$ are very similar for both forward and backward runs. Parameters $g$, $\delta$ and $f$ strongly and monotonically increases with $R$ increase the HOPG and three-layer $MoS_2$, while their dependences on $R$ are weaker for a bi-layer and the weakest for a single-layer $MoS_2$ (see blue curves in **Fig. 5**). The parameter $\varepsilon$ very weakly decreases with $R$ increase for the HOPG, single-layer and three-layer $MoS_2$ (see black curves in **Fig. 5**). The decrease of $\varepsilon(R)$ with $R$ increase is stronger for a bi-layer $MoS_2$, but starts only at $R > 1$.

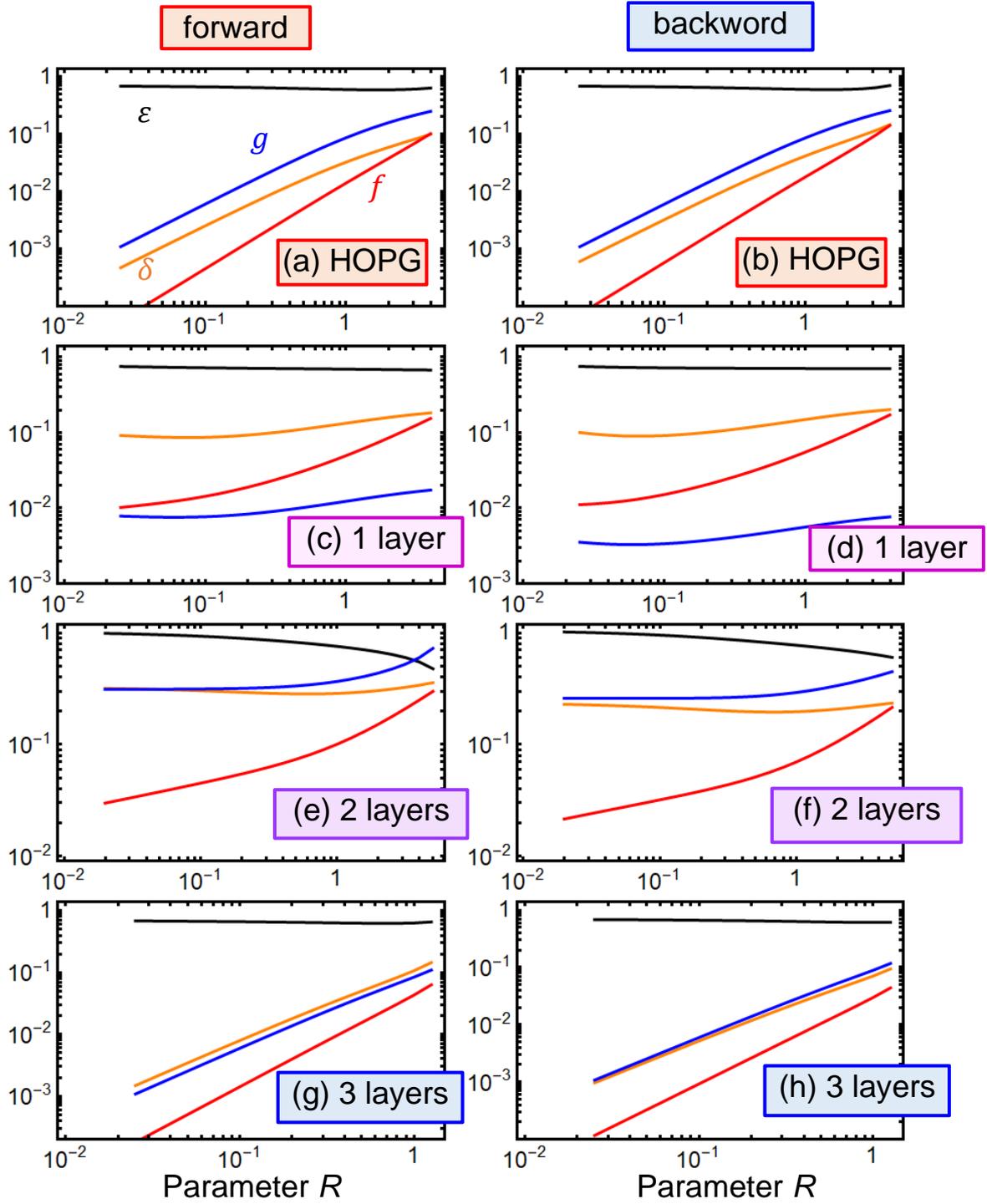

**Figure 5.** The dependence of the dimensionless fitting parameters $g$ (blue curves), $f$ (red curves), $\delta$ (orange curves) and $\varepsilon$ (black curves) on the parameter $R$ calculated for HOPG substrate (**a, b**), one (**c, d**), two (**e, f**) and three-layered (**g, h**) $MoS_2$ nanoflakes on the HOPG substrate. Plots correspond to the forward (**a, c, e, g**) and backward (**b, d, f, h**) local I-V curves of a $MoS_2$ on HOPG substrate, shown in **Fig. 3.**

Note that the dependences $g(R)$, $f(R)$, $\varepsilon(R)$ and $\delta(R)$, shown in **Fig. 5**, where obtained by numerical solution of the system of equations (8) in the range of $R$: $0 \leq R \leq 10$. It appeared that positive solutions, which have physical sense, exists in a narrower range of $R$, namely for $0.1 \leq R \leq 5$, which is shown in the figure. Since the dependences $g(R)$, $\varepsilon(R)$ and $\delta(R)$ are monotonic and unambiguous functions of $R$ the LDOS parameters can be uniquely defined for a fixed $R$. In reality the parameter $R$ is oftenly unknown due to the uncertainty of the Fermi quasi-level $E_F$ and tip-surface separation $d$. So, probabilistic methods should be used for the LDOS reconstruction.

Reconstructed LDOS for the HOPG substrate, one, two- and three-layered MoS$_2$ nanoflakes are shown in **Fig. 6** for $R = 1$. Red curves correspond to the forward direction, and blue curves correspond to the backward direction of local I-V curves, shown in **Fig. 3.** The LDOS reconstructed from the forward and backward runs are very close for the case of a HOPG substrate, single-layer and three-layer nanoflakes. The LDOS for a bi-layer nanoflake is a bit wider for a forward run.

The thinnest LDOS corresponds the HOPG substrate, the LDOS of a three-layer MoS$_2$ nanoflake is at least two times wider, the LDOS of a single-layer MoS$_2$ nanoflake is about five times wider, and the widest is the LDOS of a bi-layer MoS$_2$ nanoflake. The LDOS for the HOPG substrate and the three-layer MoS$_2$ nanoflake have a Gaussoid form with a maximum near $E = E_F$ [see **Fig. 6(a)** and **6(d)**], while the LDOS of single-layer and especially bi-layer MoS$_2$ nanoflakes consists of two Gaussoids, which are inseparable for a single-layer and separable for a bi-layer [compare **Fig. 6(b)** and **6(c)**].

The unusual form of the reconstructed LDOS can be explained by the fact that they correspond to a strongly corrugated nanoflakes, like shown in **Fig. 1(c)**. Exactly, the joint action of the size effect and corrugation can lead to the strong modulation of the free carrier density in MX$_2$ nanoflakes [4, 17]. The conclusion is corroborated by the variability and multiple jumps on the local I-V characteristics of a STM probe – sample contact (see **Fig. 3**), which reflect the randomness of a particular path of tunneling current associated with local inhomogeneity of electrical conductivity near the surface of MoS$_2$ nanoflakes, most likely caused by the changes in the nanoflake shape and their dynamic deformations during the current flow.

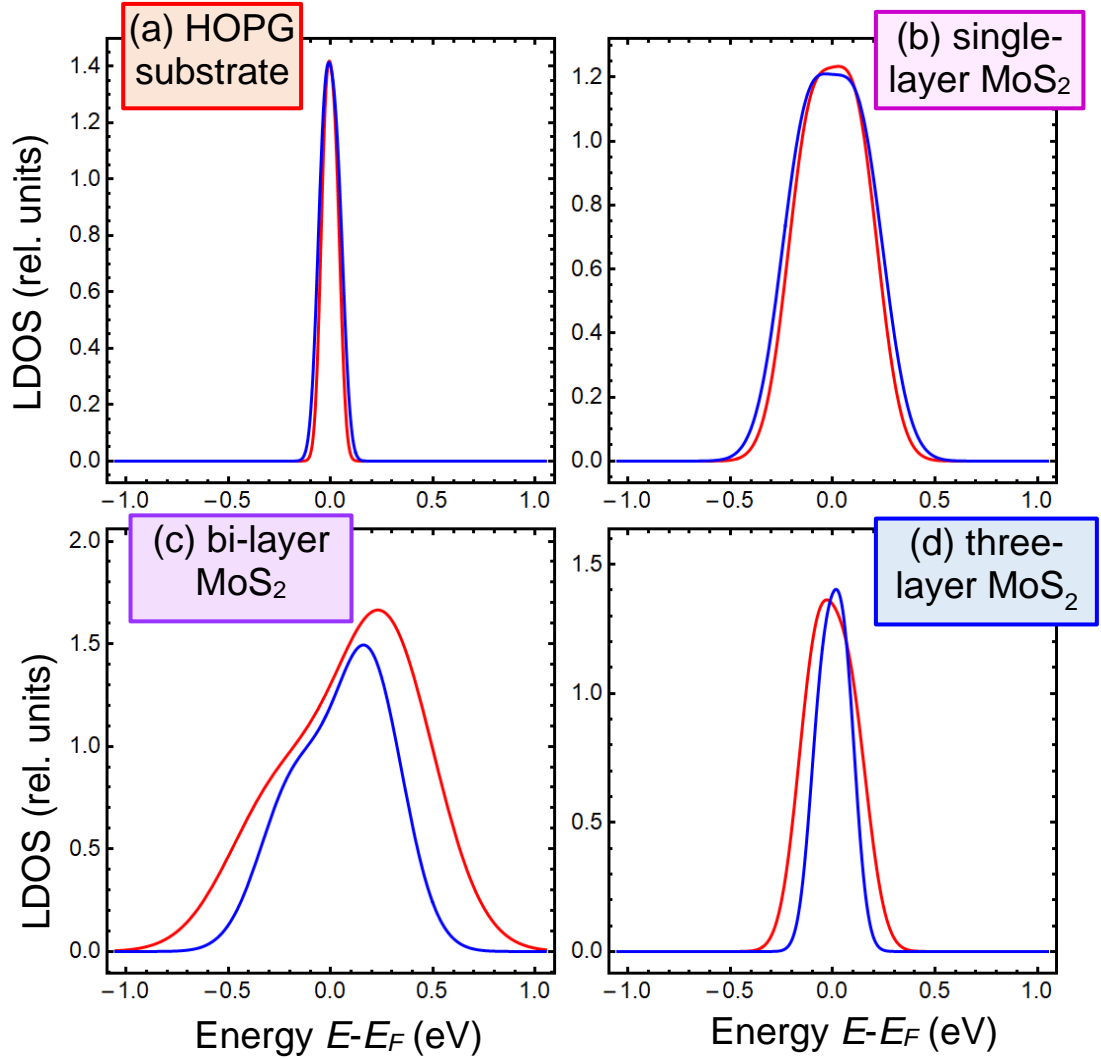

**Figure 6.** Reconstructed LDOS for HOPG substrate (**a**), one- (**b**), two- (**c**) and three-layered (**d**) MoS$_2$ nanoflakes on the substrate. Red curves correspond to the forward direction, and blue curves correspond to the backward direction of local I-V curves, shown in **Fig. 3.** Parameter $R = 1$.

## IV. Conclusions

Using a serial expansion of Tersoff formulae, we propose a flexible method how to reconstruct the LDOS from local current-voltage characteristics measured in STM experiments. We established a set of key physical parameters, which characterize the tunneling current of a STM probe – sample contact and the sample LDOS expanded in Gaussian functions. Using a direct variational method coupled with a probabilistic analysis, we determine these parameters from the STM experiments for MoS$_2$ nanoflakes with different number of layers. Three hypotheses were used, - a proper semiconductor, a semi-metal type overlapping hole and electron LDOS, and a metallic LDOS.

From the fitting of the I-V characteristics for different thickness of $MoS_2$ nanoflakes we determined the LDOS parameters, and used the interpolation function describing changes of these parameters with continuous thickening of the $MoS_2$ layer. The reconstructed LDOS reveals pronounced size effects for the single-layer, bi-layer and three-layer $MoS_2$ nanoflakes, which we relate with low dimensionality and strong bending/corrugation of the nanoflakes. The conclusion is corroborated by the variability and multiple jumps on the local I-V characteristics of a STM probe – sample contact, which reflect the randomness of a particular path of tunneling current associated with local inhomogeneity of electrical conductivity near the surface of $MoS_2$ nanoflakes, most likely caused by the changes in the nanoflake shape and their dynamic deformations during the current flow.

The main result of the work is the reconstruction of the LDOS in a relatively wide energy range around a Fermi level, which allows insight in the local band structure of LD-TMDs. We calculated it in a range of 0.5 V from the Fermi level, and reveal that the full width at half height of the LDOS is maximal for two-layer $MoS_2$, it begins to decrease for thicker nanoflakes. The unusual form of the reconstructed LDOS can be explained by the fact that they correspond to a strongly corrugated nanoflakes. Exactly, the joint action of the size effect and corrugation can lead to the strong modulation of the free carrier density in $MX_2$ nanoflakes.

The proposed elaboration of the Tersoff approach allowing LDOS reconstruction is suitable for quantitative description of versatile STM experiments, and obtained results can be useful for the microscopic physical understanding of the surface, strain and bending contribution to LD-TMDs electronic properties.

**Samples preparation and STM measurements technique.** In order to characterize the structure of $MoS_2$ nanoflakes by STM, physadsorbed $2D$-$MoS_2$ nanoflakes were deposited on the surface from a solution in n-tetradecane ($C_{14}H_{30}$). To do this, a drop of solution was applied to a freshly made substrate of highly oriented pyrolytic graphite (HOPG). Scanning of the samples was performed in non-vacuum conditions, varying the scan parameters. The probes were made by mechanical sharpening of the wire Pt-Ir (80:20) with a diameter of 0.25 mm. Typical scanning parameters: voltage $U_t = 0.5$-$1.8$ V, current $I_t = 50$-$200$ pA. Scanning speed varied from 0.1 μm/s to 3 μm/s. STM picture was obtained in the DC mode. Local I – V characteristics were measured with open feedback and locked scanning. SEM images of $MoS_2$ nanoflakes were obtained on a conductive gold substrate; and formed by secondary electrons in a Tescan Vega 3 SBH EasyProbe scanning electron microscope. The sample has the form of a thin crystal of a complex shape deformed by stresses. The thickness was about 1-3 layers, arranged randomly. For the elemental analysis, a Bruker Quantax 610M spectrometer was used, which showed the presence of Mo, S and Cu and O impurities in trace quantities.

**Authors' contribution.** A.N.M. generated the research idea, formulated the problem, performed analytical calculations (assisted by M.Ye. and E.A.E.), interpreted theoretical and experimental results, and wrote the manuscript draft. H.V.S. performed the mathematical treatment of STM results, obtained by Y.Yu.L. E.A.E. performed fitting and prepared figures. G.I.D. and M.V.O. prepared the samples and performed SEM measurements. G.S.S., Y.K. and S.V.K. worked on the manuscript improvement.

**Data Availability Statement:** Data supporting reported results was visualized in Mathematica 12.2 [31] and can be found at the Notebook Archive (https://notebookarchive.org/xxx).

**Conflicts of Interest:** The authors declare no conflict of interest.

**Acknowledgments.** A.N.M. and G.I.D. are very grateful to Prof. Victor Vainberg and Prof. Alexander Marchenko for useful discussions and valuable suggestions.

**Funding.** This work (A.N.M., H.V.S., Y.Yu.L. and G.I.D.) has been supported by the National Research Fund of Ukraine (project "Low-dimensional graphene-like transition metal dichalcogenides with controllable polar and electronic properties for advanced nanoelectronics and biomedical applications", grant application 2020.02/0027). This effort (S.V.K.) was supported by the Oak Ridge National Laboratory's Center for Nanophase Materials Sciences (CNMS), a U.S. Department of Energy, Office of Science User Facility. Y.K. research was supported by the National Research Foundation of Korea (NRF) grant funded by the Korea government(MSIT) (No. NRF-2021R1A2C2009642).

## Appendix A

The tunneling current, as Tersoff assumes, is given by the following formula:

$$I(x,y) = \frac{2\pi e}{\hbar} \sum_{\mu\nu} f(E_\mu) \left[1 - f(E_\nu + eV)\right] \left|M_{\mu\nu}(x,y)\right|^2 \delta(E_\mu - E_\nu), \quad \text{(A.1a)}$$

where $f(E) = \frac{1}{1+exp\left(\frac{E-E_F}{k_B T}\right)}$ – is the Fermi function, $E_F$ – is the Fermi level, $eV$ – current voltage, which may depend on the "transverse" relief of the surface. The matrix element $M_{\mu\nu}$ is the following integral

$$M_{\mu\nu} = -\frac{\hbar^2}{2m} \int \left(\psi_\mu^* \nabla \psi_\nu - \psi_\nu^* \nabla \psi_\mu\right) ds \qquad \text{(A.1b)}$$

Let us transform the Fermi functions. We can write as follows:

$$f(\varepsilon) = \frac{1}{1+exp(\varepsilon)}, \qquad 1 - f(\varepsilon) = f(-\varepsilon) \qquad \text{(A.2)}$$

The Taylor series formula gives the following:

$$f(E_\nu + eV) \approx f(E_\nu) + f'(E_\nu)\, eV \qquad \text{(A.2c)}$$

Thus, our formula (A.1a) can be written in the following form:

$$I(x,y) = \frac{2\pi e}{\hbar} \sum_{\mu\nu} f(E_\mu) \left[f(-E_\nu) - f'(-E_\nu)\, eV\right] \left|M_{\mu\nu}(x,y)\right|^2 \delta(E_\mu - E_\nu), \quad \text{(A.3)}$$

Assuming that we operate around the Fermi level, we can write as follows (using θ as the Heaviside function):

$$f(E) \approx \theta(E - E_F), \ \ f'(E) \approx \delta(E - E_F) \qquad \text{(A.4)}$$

Thus, we can write that

$$I(x,y) = \frac{2\pi e}{\hbar} \sum_{\mu\nu} \theta(E_\mu - E_F) \left[\theta(-E_\nu + E_F) - \delta(E_\nu - E_F)\, eV\right] \left|M_{\mu\nu}(x,y)\right|^2 \delta(E_\mu - E_\nu), \quad \text{(A.5)}$$

Or

$$I(x,y) = \frac{2\pi e}{\hbar} \sum_\nu \theta(E_\nu - E_F) \left[\theta(-E_\nu + E_F) - \delta(E_\nu - E_F)\, eV\right] |M_{\nu\nu}(x,y)|^2 \qquad \text{(A.6)}$$

It can be safely assumed that $\theta(x)\delta(x) = \frac{1}{2}, x = 0$. And $\theta(x)\theta(-x) \equiv 0$. Thus, we write

$$I(x,y) = -e^2 V \frac{\pi}{\hbar} \sum_\nu |M_{\nu\nu}(x,y)|^2 \qquad \text{(A.6)}$$

## Appendix B

Let us expand the tunneling current (4) in series either on the voltage powers:

$$I(V) \cong \sum_{i=1}^n \tilde{\sigma}_i (V - V_S)^i, \qquad \text{(B.1a)}$$

where $n$ is an integer, and $V_S$ is a voltage shift determined by $I_S$ and vise a versa. The first two expansion coefficients are:

$$\tilde{\sigma}_1 = \rho \cdot \rho_S(\boldsymbol{r}, V_S) \exp\left[-\sqrt{Z\left(\phi - \frac{V_S}{2}\right)}\right], \tag{B.1b}$$

$$\tilde{\sigma}_2 = \rho \cdot \frac{d}{dV}\left(\rho_S(\boldsymbol{r}, V) \exp\left[-\sqrt{Z\left(\phi - \frac{V}{2}\right)}\right]\right)\Bigg|_{V=V_S}. \tag{B.1c}$$

The dependence of the fitting parameters $\tilde{\sigma}_i$ and $V_S$ on the number of MoS$_2$ layers (0, 1, 2, and 3) extracted for the forward (red symbols) and backward (blue symbols) local I-V curves of a MoS$_2$ on HOPG substrate, shown in **Fig. 2,** are shown in **Fig. B1**.

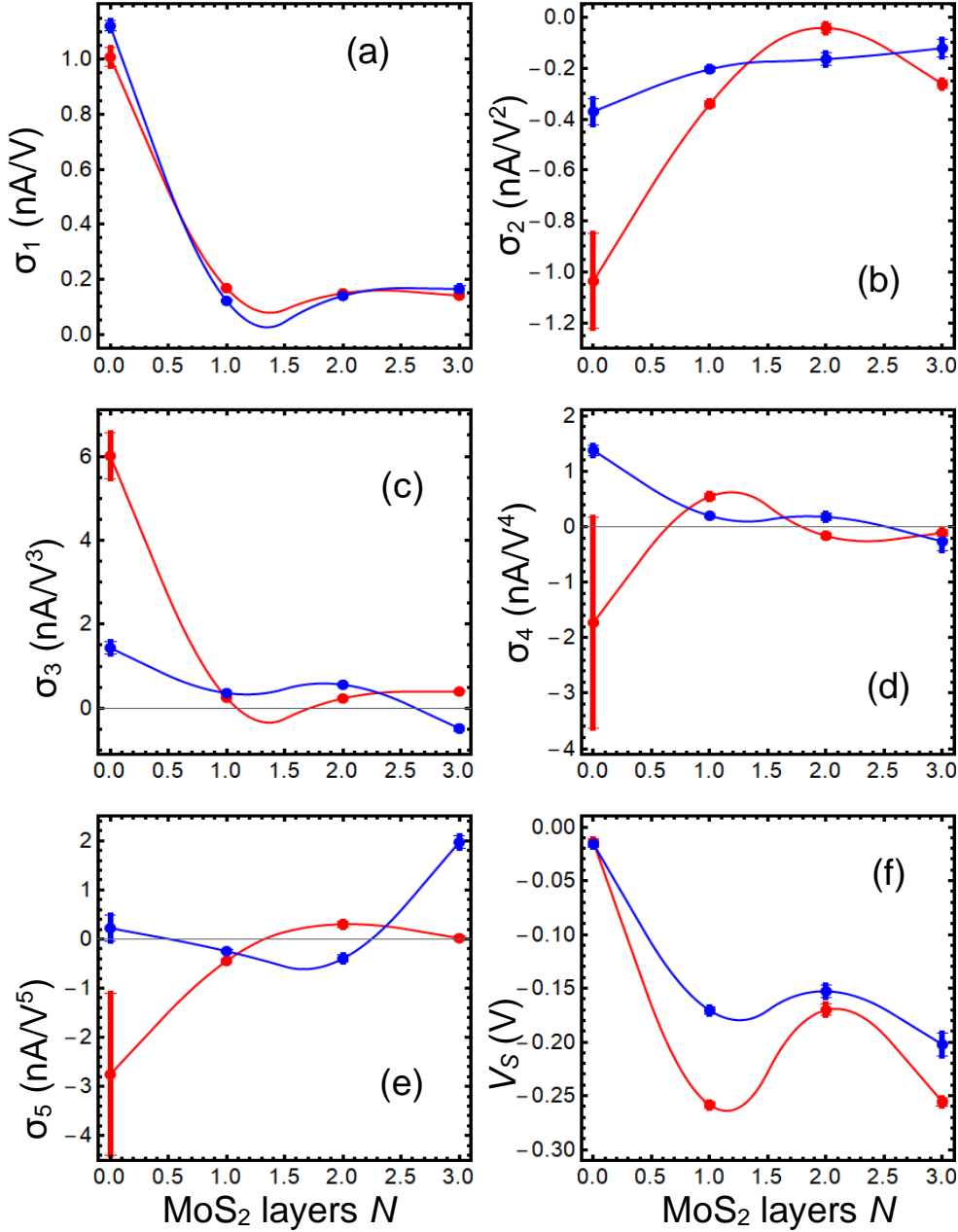

**Figure B1.** The dependence of the fitting parameters $\sigma_i$ and $V_S$ on the number $N$ of MoS$_2$ layers (0, 1, 2, and 3) extracted for the forward (red symbols) and backward (blue symbols) local I-V curves of a MoS$_2$ on HOPG substrate, shown in **Fig. 2.** Solid red and blue curves are spline-interpolations.

## Appendix C

The system of equations for the determination of LDOS parameters in Eq.(6) has the form:

$$\sigma_1 = \rho(g_e + g_p)\exp\left[-\sqrt{Z\phi} - \frac{E_s^2}{\Delta^2}\right], \tag{C.1a}$$

$$\sigma_2 = \rho(g_e - g_p)\frac{E_s}{\Delta^2}\exp\left[-\sqrt{Z\phi} - \frac{E_s^2}{\Delta^2}\right], \tag{C.1b}$$

$$\sigma_3 = \frac{\rho}{6}\exp\left[-\sqrt{Z\phi} - \frac{E_s^2}{\Delta^2}\right]\left[\frac{Z\phi+\sqrt{Z\phi}}{16\phi^2}(g_e + g_p) + \frac{1}{2}\sqrt{\frac{Z}{\phi}}(g_e - g_p)\frac{E_s}{\Delta^2} + 2(g_e + g_p)\frac{2E_s^2-\Delta^2}{\Delta^4}\right], \tag{C.1c}$$

$$\sigma_4 = \frac{\rho}{24}\exp\left[-\sqrt{Z\phi} - \frac{E_s^2}{\Delta^2}\right]\left[\frac{Z\phi+\sqrt{Z\phi}}{4\phi^2}(g_e - g_p)\frac{E_s}{\Delta^2} + \sqrt{\frac{Z}{\phi}}\frac{2E_s^2-\Delta^2}{2\Delta^4}(g_e + g_p) + \right.$$
$$\left. 4\frac{2E_s^2-3\Delta^2}{\Delta^4}\frac{E_s}{\Delta^2}(g_e - g_p)\right]. \tag{C.1d}$$

$$\sigma_5 = \frac{\rho}{120}\exp\left[-\sqrt{Z\phi} - \frac{E_s^2}{\Delta^2}\right]\left[\frac{15Z\phi+Z^2\phi^2+15\sqrt{Z\phi}+6(Z\phi)^{3/2}}{256\phi^4}(g_e + g_p) + \right.$$
$$\frac{3Z\phi+3\sqrt{Z\phi}+(Z\phi)^{3/2}}{16\phi^3}(g_e - g_p)\frac{E_s}{\Delta^2} + \frac{Z\phi+\sqrt{Z\phi}}{2\phi^2}\frac{2E_s^2-\Delta^2}{\Delta^4}(g_e + g_p) + 3\sqrt{\frac{Z}{\phi}}\frac{2E_s^2-3\Delta^2}{\Delta^4}(g_e - $$
$$\left. g_p)\frac{E_s}{\Delta^2} + 4\frac{4E_s^2(E_s^2-3\Delta^2)+3\Delta^4}{\Delta^8}(g_e + g_p)\right]. \tag{C.1e}$$

Equation (C.1) can be rewritten in the form:

$$\sigma_1 = \rho g_e(g_e + g_p)\exp\left[-\sqrt{Z\phi} - \frac{E_s^2}{\Delta^2}\right], \tag{C.2a}$$

$$\sigma_2 = \sigma_1\frac{g_e-g_p}{g_e+g_p}\frac{E_s}{\Delta^2}, \tag{C.2b}$$

$$\sigma_3 = \frac{\sigma_1}{6}\left[\frac{Z\phi+\sqrt{Z\phi}}{16\phi^2} + \frac{1}{2}\sqrt{\frac{Z}{\phi}}\frac{g_e-g_p}{g_e+g_p}\frac{E_s}{\Delta^2} + 2\frac{2E_s^2-\Delta^2}{\Delta^4}\right], \tag{C.2c}$$

$$\sigma_4 = \frac{\sigma_1}{24}\left[\frac{Z\phi+\sqrt{Z\phi}}{4\phi^2}\frac{g_e-g_p}{g_e+g_p}\frac{E_s}{\Delta^2} + \sqrt{\frac{Z}{\phi}}\frac{2E_s^2-\Delta^2}{2\Delta^4} + 4\frac{2E_s^2-3\Delta^2}{\Delta^4}\frac{E_s}{\Delta^2}\frac{g_e-g_p}{g_e+g_p}\right], \tag{C.2d}$$

$$\sigma_5 = \frac{\sigma_1}{120}\left[\frac{15Z\phi+Z^2\phi^2+15\sqrt{Z\phi}+6(Z\phi)^{3/2}}{256\phi^4} + \frac{3Z\phi+3\sqrt{Z\phi}+(Z\phi)^{3/2}}{16\phi^3}\frac{g_e-g_p}{g_e+g_p}\frac{E_s}{\Delta^2} + \frac{Z\phi+\sqrt{Z\phi}}{2\phi^2}\frac{2E_s^2-\Delta^2}{\Delta^4} + \right.$$
$$\left. 3\sqrt{\frac{Z}{\phi}}\frac{2E_s^2-3\Delta^2}{\Delta^4}\frac{g_e-g_p}{g_e+g_p}\frac{E_s}{\Delta^2} + 4\frac{4E_s^2(E_s^2-3\Delta^2)+3\Delta^4}{\Delta^8}\right]. \tag{C.2e}$$

Below we introduce dimensionless variables

$$R = \sqrt{Z\phi}, \quad f = \frac{\Delta}{|\phi|}, \quad \varepsilon = \frac{E_s}{\Delta}, \quad \delta = \frac{\Delta}{E_g}, \quad g = \frac{g_e-g_p}{g_e+g_p} \tag{C.3}$$

Where the dispersion $\Delta$ is dimensionalized on the bulk band gap of MoS$_2$, $E_g = 1.2$eV. Using Eqs.(C.2) and (C.3) one could get the following system of equations for the dimensionless variables $\delta$, $f$, $R$ and $\varepsilon$:

$$\Delta \frac{\sigma_2}{\sigma_1} = E_g \frac{\sigma_2}{\sigma_1} \delta = g\varepsilon, \tag{C.4a}$$

$$6\Delta^2 \frac{\sigma_3}{\sigma_1} = 6E_g^2 \frac{\sigma_3}{\sigma_1} \delta^2 = \frac{R^2+R}{16} f^2 + \frac{f}{2} Rg\varepsilon + 2(2\varepsilon^2 - 1), \tag{C.4b}$$

$$24 \frac{\sigma_4}{\sigma_1} \Delta^3 = 24E_g^3 \frac{\sigma_4}{\sigma_1} \delta^3 = \frac{R^2+R}{4} f^2 g\varepsilon + fR \left( \varepsilon^2 - \frac{1}{2} \right) + 4(2\varepsilon^2 - 3)g\varepsilon, \tag{C.4c}$$

$$120 \frac{\sigma_5}{\sigma_1} \Delta^4 = 120E_g^4 \frac{\sigma_5}{\sigma_1} \delta^4 = f^4 \frac{15(R+R^2)+6R^3+R^4}{256} + f^3 \frac{3(R+R^2)+R^3}{16} g\varepsilon + f^2(R^2 + R) \left( \varepsilon^2 - \frac{1}{2} \right) + 3fR(2\varepsilon^2 - 3)g\varepsilon + 4(4\varepsilon^4 - 12\varepsilon^2 + 3). \tag{C.4d}$$

The system (C.4) can be solved numerically. A practically important case $\sigma_2 = \sigma_4 = 0$ (corresponding to the experimental data for an antisymmetric I-V curve) should be considered separately, and in this case we obtain much simpler system of equations:

$$\sigma_2 = \sigma_4 = 0 \implies g = 0, \ \varepsilon^2 = \frac{1}{2}, \tag{C.5a}$$

$$6E_g^2 \frac{\sigma_3}{\sigma_1} \delta^2 = \frac{R^2+R}{16} f^2, \tag{C.5b}$$

$$120E_g^4 \frac{\sigma_5}{\sigma_1} \delta^4 = f^4 \frac{15(R+R^2)+6R^3+R^4}{256} + 4(4\varepsilon^4 - 12\varepsilon^2 + 3). \tag{C.5c}$$

The description of the tip and the sample LDOS is given in **Table CI.** Note that some of the sample LDOS parameters can be determined only in combination with the tip LDOS parameters.

**Table CI.** Description of the tip and the sample LDOS

| N | Parameter | Definition | | Comments |
|---|---|---|---|---|
| 1 | $\rho$ | $\frac{4\pi}{h} \rho_T$ | global | the parameter is identical for all $\sigma_i^{exp}$, because it is related with STM tip DOS |
| 2 | $Z$ | $\frac{8em}{\hbar^2} d^2$ | global | the parameter is identical for all $\sigma_i^{exp}$, because it is related with the carrier mass in the gap and gap width |
| 3 | $\phi$ | $\frac{\varphi_T + \varphi_S}{2} - E_F$ | local | $\varphi_S$ depends on the sample LDOS |
| 4 | $g_i$ | $i = e, p$ | local | number of electron/hole states in the sample LDOS |
| 5 | $E_i$ | $i = e, p$ | local | characteristic energy in the sample LDOS |
| 6 | $\Delta_i$ | $i = e, p$ | local | Dispersion of the sample LDOS |
| 7 | $E_F$ | Fermi level | local | Position of the system Fermi level depends on the tip and sample LDOS |